\documentclass[12pt]{article}

\usepackage{hyperref}
\hypersetup{
    colorlinks=true, linkcolor=black, filecolor=red, urlcolor=blue, 
    pdftitle={Overleaf Example}
    pdfpagemode=FullScreen,
    }
\usepackage{url}

\usepackage{amssymb}
\usepackage{graphicx}        
\usepackage{makeidx}
\usepackage{epsf}

\makeindex

\def\fnote#1#2{\begingroup\def\thefootnote{#1}\footnote{#2}\addtocounter{footnote}{-1}\endgroup}

\def\a{\alpha}   \def\b{\beta}    \def\g{\gamma}  
\def\e{\epsilon} \def\G{\Gamma}     
\def\L{\Lambda}     \def\si{\sigma}

 \def\cH{{\cal H}}

\def\cP{{\cal P}}  \def\cR{{\cal R}}

   \def\bE{{\bar E}}  \def\bF{{\bar F}}

\def\Ewhat{{\widehat E}}

\def\adot{{\dot{a}}}

            \def\rmIm{{\rm Im}}

\def\rmPl{{\rm Pl}}            

       \def\rmSL{{\rm SL}}      \def\rmSO{{\rm SO}}

            \def\rmmod{{\rm mod}}

\newcommand{\rmsw}{{\rm sw}}

\def\rmRe{{\rm Re}}     

 \def\mathC{{\mathbb C}} 
  
  \def\mathR{{\mathbb R}}
\def\mathZ{{\mathbb Z}}

\def\fnote#1#2{\begingroup\def\thefootnote{#1}\footnote{#2}\addtocounter{footnote}{-1}\endgroup}

\def\beq{\begin{equation}}
\def\eeq{\end{equation}}
\def\bea{\begin{eqnarray}}
\def\eea{\end{eqnarray}}

\def\lleq#1{\label{#1}\eeq}
\let\nn=\nonumber

\def\notin{\ \hbox{{$\in$}\kern-.51em\hbox{/}}}

  \def\E1Fq{E_1/\IF_q}

\def\notdiv{{\relax{~|\kern-.34em /~}}}

 \def\otau{{\overline{\tau}}}

\def\boxit#1{
\vbox{\hrule height1pt\hbox{\vrule width1pt\kern0.3cm
\vbox{\kern0.3cm\hbox{$\displaystyle#1$}\kern0.3cm}\kern0.3cm\vrule
width1pt}\hrule height1pt}}

\begin{document}

\baselineskip=16pt

\hfill \phantom{\bf Draft}\break
\phantom{w} 
\hfill \today

\parindent=0pt

\vskip .7truein

\centerline{\Large {\bf Scaling Behavior of Observables as a} }
\vskip .1truein

\centerline{\Large {\bf  Model Characteristic in Multifield Inflation}}

\vskip .4truein

 \centerline{ {\sc~ Monika Lynker\fnote{$\star$}{mlynker@iu.edu} and 
                           Rolf Schimmrigk\fnote{$\diamond$}{netahu@yahoo.com, rschimmr@iu.edu}}}

\vskip .2truein

\centerline{Dept. of Physics}
\centerline{Indiana University South Bend} 
\centerline{1700 Mishawaka Ave., South Bend, IN 46615}

\vskip .8truein

\centerline{\bf Abstract} 
\vskip .1truein


 One of the fundamental questions in inflation is how to characterize the structure of different types of models  in the 
 field theoretic landscape.  Proposals in this direction include attempts to directly characterize the formal structure 
 of the theory by considering complexity measures of  the potentials.
 An alternative intrinsic approach is to focus on the behavior of the observables that result 
 from different models and to ask whether their behavior differs among  models. This 
 type of analysis can be applied even to nontrivial multifield theories where a natural measure 
 of the complexity of the model is not obvious and the analytical evaluation of the observables is often impossible. 
 In such cases one may still compute these observables numerically and investigate their behavior.
 One interesting case is when observables show a scaling behavior, in which case  theories can be characterized 
 in terms of their scaling amplitudes and exponents. Generically, models have nontrivial 
 parameter spaces, leading to exponents that are functions of these parameters. In such cases we consider 
 an iterative procedure to determine whether the exponent functions in turn lead to a scaling behavior.
 We show that modular inflation models can be characterized by families of simple scaling laws
 and that the scaling exponents that arise in this way in turn show scaling in dependence of the varying 
 energy scales.

\renewcommand\thepage{}
\newpage
\parindent=0pt

\pagenumbering{arabic}

\vfill \eject

\baselineskip=17pt
\parskip=0.01truein

\tableofcontents

\vskip .5truein

\baselineskip=21pt
\parskip=.15truein

\section{Introduction} 

One of the key questions in inflation is how to characterize  different models in a systematic way, and  in the 
process define classes of models.  In the past proposals in this direction have been made that involved quite 
different strategies.  Often the focus has been on the formal structure of the potentials, attempting to formulate measures
 of complexity, either through the algebraic structure of the potentials, or through the number of parameters involved 
 (see for example ref. \cite{bst05} for an early excursion and ref. \cite{b08etal} for more discussion).
 The rationale of algebraic simplicity is tempting, but goes counter to what has been learned about the structure 
 of physical theories
 in the past century.  A more common  point of view is to focus on the conceptual structure 
 of a theory, 
 which may lead to simplicity in terms of 
 ideas even though the algebraic  structure might be complicated. To apply this thinking to inflationary models requires 
 their embedding
 into a broader framework, such as string theory. The price to pay in this case is a large theoretical structure that 
 in the end may have little to 
 do with the inflationary model under consideration \cite{b13etal, bm14}. A third point of view aims to make 
 the conceptual picture more concrete by asking whether symmetries motivated by fundamental issues in inflation
  might play a role (see e.g.  \cite{rs14, rs15, k20} for an illustration).
 
The advantage of the inflationary framework is that it is constrained by data. The most stringent of these constraints are
currently provided by the {\sc Planck} probe in combination with the BICEP/Keck experiment
\cite{pbk21}, although other ground based observatories, already operating or coming on line in the near future, will 
 improve these  results \cite{class21, simons21, cmbs4}. 
It is therefore natural to look at the output of the theory instead of the input by analyzing the behavior of the observables in the 
range of the parameter space that is phenomenologically viable. 
In this point of view the goal is to detect structure among the parameters and to 
distinguish different models by the types of their observable behavior.   Such an approach can be guided by 
 simple models in which 
 the analytic structure of the observables can, within certain approximations, be derived in an analytic form. Paradigmatic 
 examples that have been considered in  a large number of papers include monomial 
inflation \cite{l83, t83}, the scalar field version of the Starobinsky model  \cite{s80} and 
hilltop models \cite{km95, l96, bl05}, among others.
In these cases the  spectral index and the tensor-to-scalar ratio of the Lukash-Bardeen 
perturbation $\cR$ can be shown to scale in terms of the number of e-folds $N$. 

Multifield inflation models are in general too complicated to be solved analytically and as  a result the behavior 
of their observables in dependence of other parameters cannot be described in closed form. Nevertheless, these models can be analyzed numerically
  and possible correlations can be explored. In this paper we propose to explore the problem of 
  characterizing multifield inflation models
   by considering the scaling behavior of their observables.   If such a behavior can be established the possibility arises 
that multifield inflation models fall into different classes that allow to characterize  different regions in 
the inflationary landscape 
by the type of their scaling behavior, as well as a quantitative foliation of the theory space in terms of their 
 scaling structure. This provides a possible procedure to identify types of multifield models that extends beyond 
 the symmetry based strategy considered in \cite{rs14, rs15}.

 Inflationary models will in general have parameters beyond the overall energy scale. In those cases
  a scaling relation between the observables and the number of e-folds will lead to a family of relations with 
  amplitudes and exponents that depend on these additional parameters, generically denoted  here
  by $\mu$.  Given such a family of exponents $\b(\mu)$  we can  ask whether there exists in turn a scaling 
  relation for this family 
  of scaling exponents. In twofield models with two energy scales this is the point 
where the iteration stops and if scaling exists in this second iteration there will be a single scaling exponent, 
obtained from the $\b(\mu)$, that now can be used to characterize the model.
In this paper we will apply the procedure just outlined to modular inflation models, in 
particular $j$-inflation \cite{rs14, rs16, rs21} 
and $h_2$-inflation \cite{ls19}. These theories are characterized by energy scales $(\L,\mu)$  
and we will show that there exists a family of scaling exponents $\b(\mu)$ which in turn lead to a further exponent.
We will show that modular inflation restricted to a {\sc Planck} compatible 
range of parameters leads to a scaling regime that differs from that of singlefield inflation classes. 
It would be interesting to see whether other multifield models exhibit a 
similar scaling behavior, and if so what the exponents are in comparison.

This paper is organized as follows. 
In Section 2 we briefly discuss the scaling behavior of some singlefield inflation models that provide basic reference
        classes to compare with multifield inflation behavior.
In Section 3 we briefly outline the structure of modular inflation as the class of models we will focus on. 
In Section 4 we analyze in detail the behavior of two modular inflation models and establish their scaling behavior.
In Section 5 we analyze the scaling of the scaling exponents as the fundamental energy scale is varied in the models.
In Section 6 we present our conclusions.

\vskip 0.3truein

\section{Scaling in singlefield inflation}

In this section we briefly discuss the behavior encountered in some simple singlefield inflation models, which will 
allow us to compare our results with previous exponents, obtained by considering approximations strong enough to 
admit analytical considerations. 
The focus of this paper is on the behavior of the observables associated to the 
 Lukash-Bardeen perturbation  \cite{l80, b80}
\beq
 \cR ~=~ H \delta u ~-~ \psi,
 \eeq
 where $H=\adot/a$ is the Hubble-Slipher parameter, $\psi$ is the metric potential, and $\delta u$ is 
 obtained from $\delta T_{0i}$,  as detailed for example in \cite{w08, rs16}.
 In the slow-roll approximation of singlefield inflation 
 the observables are determined in terms of the slow-roll 
 parameters $\e_V$ and $\eta_V$ defined as
    \beq
    \e_V ~=~ \frac{M_\rmPl^2}{2} \left(\frac{V'}{V}\right)^2, ~~~~~
    \eta_V ~=~ M_\rmPl^2 \frac{V''}{V}.
   \eeq
 The spectral index $n_{\cR\cR}$ and the tensor ratio $r$ can then can be written as
\beq
 n_{\cR\cR} = 1 - 6\e_V + 2\eta_V,~~~~~  r ~=~ 16\e_V.
  \eeq
 The third parameter relevant in our discussion is the number of e-folds $N_*$ 
\beq
 N_* ~=~ \int_{t_*}^{t_e} H(t) dt.
 \eeq
 
In some simple theories it is  possible to invert the $N_*=N(\phi_*)$ relation, given appropriate approximations. 
 This allows to express both the spectral index and the tensor-to-scalar ratio analytically as functions of $N_*$, leading 
to a scaling behavior of the observables. The earliest paradigmatic classes can be found in  monomial inflation \cite{l83, t83} 
and the  Starobinsky model \cite{s80}, leading however to  very restricted scaling types. More general types are 
 provided by hilltop and inverse hilltop models. In a somewhat different context the $1/N$ parameter was suggested 
 in early work in the framework of singlefield inflation as an expansion 
parameter \cite{m95etal, b05etal} and later in \cite{m13, r13}. 
We  will  discuss a few  of these models briefly  because they suggest natural questions that 
can be raised  in the more general context of multifield inflation.

\underline{Monomial inflation}

In singlefield monomial inflation with potentials $V_p = \L^{4-p} \phi^p$ \cite{l83, t83}
  one finds in the slow-roll approximation, and the approximation that the horizon crossing value $\phi_*$ is much 
  larger than the inflaton value $\phi_e$ at the end of inflation, the scaling 
  behavior
\beq
 n_{\cR\cR}(N) ~=~ 1 - \frac{\a_p}{N_*},~~~~~ r(N) ~=~ \frac{\b_p}{N_*},
 \eeq
 where $\a_p,\b_p$ are constants for fixed $p$. 
 This behavior comes about because the slow-roll parameters $\e_V$ and $\eta_V$ are 
 proportional for  arbitrary $p$ and $\e_V$ scales like $\e_V \cong  N^{-1}$.
 This  has led to the exclusion of this class of models by the ever more decreasing experimental bounds on 
 the tensor ratio $r$ over the past decade \cite{pbk21}.  Other singlefield models with a scaling behavior of monomial type 
 exist and the totality of such models form what one could call
 the monomial scaling class for the observables $(n_{\cR\cR}, r)$.
 
 \underline{Starobinsky inflation}
  
  A further prominent singlefield inflation model is Starobinsky inflation \cite{s80},  which is usually 
  considered in its potential form as
  \beq
   V ~=~ \L^4 \left(1- e^{-\sqrt{\frac{2}{3}} \frac{\phi}{M_\rmPl}} \right)^2.
  \eeq
  An analytical form of scaling can be obtained by adopting the slow-roll and the large field approximation, 
  which allows to write a scaling behavior of the type
  \beq
   n_{\cR\cR}(N) ~=~ 1 - \frac{\a_\rmsw}{N_*},~~~~~  r(N) ~=~ \frac{\b_\rmsw}{N_*^2},
   \eeq
   where $\a_\rmsw$ and $\b_\rmsw$ are constants.  This arises because the two slow-roll parameters of this model
    have different scaling behavior as 
   \beq
   \e_V \cong N_*^{-2}, ~~~~~\eta_V \cong -N_*^{-1}.
   \eeq
   Hence the $\e_V$ term is suppressed in the spectral index, which is essentially determined by $\eta_V$,
   and  the decay of the tensor-to-scalar ratio with $N_*$ is stronger in Starobinsky inflation than in monomial inflation.
   As noted above, it is because of this different scaling behavior that the Starobinsky model remains viable.
   Other models that survive for similar reasons 
   exist in this type of model, defining the Starobinsky type scaling class  for the observables $n_{\cR\cR}$ and $r$.

\underline{Other singlefield scaling classes}

A family of models that leads to a whole sequence of different scaling relations is  hilltop inflation, defined as
\beq
 V_h ~=~ \L^4\left(1 - \left(\frac{\phi}{\mu}\right)^p\right),
 \eeq
 where $p$ is a priori undetermined. It was for example pointed out in refs. \cite{km95, l96} that for $p\neq 2$ 
 appropriate approximations lead to analytical power scaling behavior of the  form
   \beq
 n_{\cR\cR} ~=~ 1 - \frac{2(p-1)}{(p-2)} \frac{1}{N}, ~~~~~~~r(N) ~=~ \frac{\a_p}{N^{\b_p}}, 
 \eeq
where the tensor exponents $\b_p$ are given by
 \beq
 \b_p ~=~ \frac{2(p-1)}{p-2}.
 \eeq
 For $p=2$ the scaling behavior is exponential. This sequence of models would seem 
 to dramatically extend the range of the power scaling exponents $\b_p$ since these exponents grow as $p$ approaches two. 
 However models in this limit are not compatible with {\sc Planck} data. The exponential scaling for the $p=2$ model furthermore 
 relies on the small field approximation, which is inconsistent for this model.

 \vskip .3truein
 
\section{Modular inflation}

In this section we briefly recall the models we will analyze. More details for the
general framework can be found in the refs. \cite{rs14, rs15} and the phenomenology of these models has been analyzed in 
\cite{rs16, rs17, ls19, rs21}, with related work in \cite{pv18}. In the context of high energy inflation it is of interest to 
consider classes of models in which the symmetries of the free theory are not completely 
broken by the potential. If the remaining symmetry groups are of a certain type then such potentials can describe both large and small 
field inflation at the same time because viable islands in the field space are mapped into each other by the symmetry \cite{rs21}.
Furthermore, implications for the swampland have been discussed in \cite{rs18, ls19}.

Modular inflation is a framework for twofield inflation models that specializes the more general context of automorphic 
inflation introduced in \cite{rs14, rs15} for an arbitrary number of fields to two fields. 
The general idea is to consider field spaces spanned by coordinates $\phi^I$ that are 
defined by homogeneous spaces of the type $G/K$, where $G$ is a reductive group and $K$ is a compact subgroup. 
The target space inherits a curved metric $G_{IJ}(\phi^K)$ that is essentially induced by the group $G$.   
The inflaton potentials $V(\phi^I)$ are assumed to be invariant under the discrete group $G(\mathZ)$ obtained by restricting 
the continuous group $G(\mathR)$  to integers $\mathZ$. Such potentials can be obtained via 
 automorphic functions constructed from automorphic forms.

In the special context of modular inflation the group $G$ can be chosen as the M\"obius group $\rmSL(2,\mathR)$ and the
 compact subgroup as $K=\rmSO(2,\mathR)$.  The resulting quotient space leads to the upper halfplane $\cH \subset \mathC$
 and the inflaton multiplet $\phi^I$ defines a sigma model with $\cH$ as its target space, leading to a class of curved multifield
  theories with field space metric given by the hyperbolic metric 
  \beq
  ds^2 ~=~ \frac{d\tau d\otau}{(\rmIm~\tau)^2}
  \eeq
  where  $\tau= (\phi^1+i\phi^2)/\mu$ is the complexified dimensionless inflaton.
 The inflaton potentials $V(\phi^I)$  are determined by modular invariant 
 functions $F(\tau)$ on the upper halfplane $\cH$
\beq
 V_n(\phi^I) ~=~ \L^4 |F(\tau)|^{2n},
 \lleq{mi-pots}
 where $n$ could be chosen as any positive integer. In the present paper we consider $n=1$. 
 The functions $F(\tau)$ can be constructed as quotients of modular forms with higher weight relative to 
 some discontinuous subgroup $\G$ of the modular group. 
  
 For discrete groups $\G$ of congruence type, introduced by Klein in his ``Stufentheorie", 
  the theories are characterized by the level $N$ of these groups, denoted by  $\G(N)$,
   and the modular functions $F(\tau)$ are 
 constructed relative to $\G(N)$. 
 The models considered in the present paper can in particular be constructed in terms of the Eisenstein 
 series $E_w(\tau)$ of weight $w$, defined as
 \beq
  E_w(\tau) ~=~ 1 ~-~\frac{2w}{B_w}  \sum_{n=1}^\infty \si_{w-1}(n) q^n
 \lleq{eisenstein}
 where $q=e^{2\pi i\tau}$, $B_w$ are the Bernoulli numbers, and  $\si_w(n)$ is the divisor function. While the $E_w$
 for $w>2$ are modular forms relative to  the full modular group, they can be used to construct models at higher level. 
 More details about these definitions 
 and the general framework can be 
 found in \cite{rs14, rs16, ls19}.  It was furthermore shown in \cite{rs21} in the context of $j$-inflation that different 
 regions in the target space lead to small or large 
 inflaton values, indicating that the dichotomy of large vs. small field inflation, often used in singlefield inflation as a 
 motivation for target values for the 
 tensor ratio, does not persist in multifield inflation as a model characteristic.  
 Reheating after $j$-inflation was discussed in ref.  \cite{rs17, pv18}.

 \subsection{Observables of modular inflation}
 
In the framework of modular inflation as formulated in \cite{rs14, rs16, ls19, rs21}  the observables associated 
to the perturbation $\cR$ can be expressed in terms of the defining modular function $F$ and its derivatives.
The spectral index takes for arbitrary $F$ the form 
\beq
   n_{\cR\cR}
   ~=~ 1 -  4 \frac{M_\rmPl^2}{\mu^2}  (\rmIm~\tau)^2 
    \left[ 2 \left|\frac{F'}{F}\right|^2
     ~-~ \rmRe \left(\frac{F''}{F'}  \cdot \frac{\bF'}{\bF}\right)_\rmmod
     ~ + ~\frac{\pi}{3} \rmIm\left(\Ewhat_2 \frac{\bF'}{\bF}\right)\right],
  \lleq{nRRsr-modular}
  where $(\cdot)_\rmmod$ selects the modular part of the quotient and 
   the quasimodular form $\Ewhat_2$ of weight two, defined as
  \beq
   \Ewhat_2(\tau) ~=~ E_2(\tau) ~-~ \frac{3}{\pi \rmIm~\tau},
  \eeq
  is obtained by combining the non-modular part of the quotient $F''/F'$ with a term that arises from the non-trivial 
  metric of the field space.
 The tensor-to-scalar ratio $r=\cP_T/\cP_{\cR\cR}$ is given in general modular inflation by
 \beq
 r ~=~ 32 \frac{M_\rmPl^2}{\mu^2} (\rmIm~\tau)^2 \left|\frac{F'}{F}\right|^2,
 \eeq
 while the number of $e$-folds takes the form 
 \beq
  N_* ~= ~ \frac{1}{\sqrt{3}} \frac{\L^2}{M_\rmPl} \int_{t_*}^{t_e}dt~ |F|.
  \eeq
  
 The above results for the spectral index and the tensor ration show that in order to develop the theory of 
 modular inflation further it is necessary to specify the first and second derivatives of the modular function 
  $F(\tau^I)$ that enters the potential. This is a nontrivial task in general, as described for the full modular group in \cite{rs16} and for 
  higher level theories in \cite{ls19}. While there is no general theory available that gives explicit expressions
   for these derivatives leading to an evaluation of the observables, this 
  problem can be solved for special classes of potentials. In the present paper we will illustrate our idea of characterizing inflationary 
  theories by their energy dependent scaling exponents with two models, introduced in \cite{rs14} and \cite{ls19} respectively.
  
  \subsection{Observables in $j-$Inflation and $h_2-$inflation}

 For $j$-inflation the potential is defined in terms of the absolute modular invariant $j$
 \beq
  V ~=~ \L^4 |j(\tau)|^2,
  \eeq
  where the absolute modular invariant $j(\tau)$,  considered by Hermite and Dedekind, but often called the Klein invariant, 
   is most conveniently written in terms of the Eisenstein series $E_4$ and the Ramanujan modular 
  form $\Delta$ as 
  \beq
  j = \frac{E_4^3}{\Delta},
  \eeq
  where $ \Delta$ is  a weight twelve form that is often defined in terms of the Dedekind $\eta$ function (see below),
   but that  can also be expressed  in terms of the Eisenstein series of weight four and six as
   \beq
    \Delta ~=~ \frac{E_4^3-E_6^2}{1728}.
    \eeq
 The observables specialize to
 \beq
 n_{\cR\cR}^j ~=~ 1 - \frac{8\pi^2}{3} (\rmIm~\tau)^2 \frac{M_\rmPl^2}{\mu^2}
       \left[  8 \left|\frac{E_6}{E_4}\right|^2
             - 3 \rmRe \left( \frac{E_4^2}{E_6}\frac{\bE_6}{\bE_4}\right)
                  +  \rmRe\left(\Ewhat_2\frac{\bE_6}{\bE_4}\right)
      \right]
 \nn \\
\lleq{scalar-spectral-index-ahmi}
 and
  \beq
    r^j ~=~ 128 \pi^2 ~(\rmIm~\tau)^2 ~\frac{M_\rmPl^2}{\mu^2}
       ~\left|\frac{E_6}{E_4}\right|^2.
 \eeq
 This is discussed in more detail in the refs. \cite{rs14, rs16, rs21}.
 
 For the class of $h_N$-inflation the  potentials take the form 
  \beq
   V_N ~=~ \L^4 |h_N(\tau)|^2,
   \eeq
   in terms of hauptmodul functions given by the $\eta-$function quotients
      \beq
   h_N(\tau) ~=~ \left(\frac{\eta(\tau)}{\eta(N(\tau)}\right)^{24/(N-1)},
   \eeq
   where the Dedekind eta function is defined as $\eta =q^{1/24} \prod_n(1-q^n)$ with $q=e^{2\pi i \tau}$.  
   These functions $h_N$ can also be expressed in terms of the Eisenstein series 
   defined in eq. (\ref{eisenstein}), which is useful for the analysis.
  The observables were determined in \cite{ls19} to be given by 
 \bea
  n^{h_N}_{\cR\cR} &=&   1 -  \frac{4\pi^2}{3(N-1)} \frac{M_\rmPl^2}{\mu^2} (\rmIm~ \tau)^2 
        \left[\frac{(N+11)}{(N-1)}  |E_{2,N}|^2  
       -  \rmRe\left( \left( \frac{E_{4,N}}{E_{2,N}} ~-~ 2 \Ewhat_2 \right) \bE_{2,N} \right)  \right],    \nn \\
  r^{h_N} &=& \frac{128\pi^2}{(N-1)^2}   ~(\rmIm~\tau)^2 ~\frac{M_\rmPl^2}{\mu^2} |E_{2,N}|^2,
  \eea
where 
\beq
 E_{2,N}(\tau) ~=~ N E_2(N\tau) ~-~ E_2(\tau).
 \eeq
 
In these relations we have used that the tensor-to-scalar ratio $r$ in multifield inflation is given as $r=16\e_V$ with
\beq
 \e_V ~=~ \frac{1}{2}G^{IJ} \e_I\e_J, ~~~~\e_I ~=~ M_\rmPl \frac{V_{,I}}{V},
 \eeq
 where $G^{IJ}$ is the inverse of the field space metric.
 It follows from the CMB bounds on $r$ obtained  by the WMAP and {\sc Planck} collaborations 
  \cite{wmap03, planck18} that the local 
neighborhoods in the target space where viable initial values for the inflaton occur are saddle point regions. 
The concrete forms of 
these neighborhoods are locations in which $E_6$ is small in $j$-inflation and $E_{2,N}$ is small in $h_N$-inflation.

\vskip .3truein

\section{Scaling behavior in modular inflation}

In the previous section we have seen that modular inflation  theories are formulated in a conceptually simple 
framework but that the potentials are not particularly simple algebraically, a feature these models share with 
many theories. As a result the observables 
$n_{\cR\cR}$ and $r$ are not particularly simple either,  given by 
 quotients of modular forms and their derivatives. The ingredients are infinite series of the inflaton field leading 
 to  somewhat involved functions whose analytic structure is not explicit. The virtue of modularity of course is that 
 a finite number of terms suffice to determine the structure of the modular forms completely.
Because of these involved relations the e-fold equation in these models cannot be inverted and it is a priori not clear 
whether a simple functional relationship exists between the different parameters. 
In modular inflation the basic parameters of the models are given by two energy scales $(\L, \mu)$, as indicated in eq. (\ref{mi-pots}). 
The overall amplitude $\L$ is determined as usual by  the CMB amplitude measured by the satellite experiments. 
The energy scale $\mu$ is not directly determined by the power spectrum amplitude but  it is constrained by the experimental data, 
in particular by the tensor-to-scalar ratio $r$. In ref. \cite{rs21} it was shown 
that this ratio sweeps out an approximately horizontal band in the ($n_{\cR\cR}, r)$ plane used by the CMB collaborations to
describe  their confidence regions for  the gravitational contribution to the microwave background. 
As $\mu$ varies, this band moves up and down in this $(n_{\cR\cR},r)$-plane. 
It is in this way that the energy scale $\mu$ is constrained by the tensor ratio.

\subsection{Scaling behavior of the spectral index}

A natural question is whether the scaling of the spectral index encountered in the singlefield inflation models discussed above 
appears in some form also in multifield inflationary models. This can be tested for fixed scale $\mu$ by computing the function 
\beq
 f_\a(n_{\cR\cR}, N) ~=~ n_{\cR\cR} ~-~ \left(1 - \frac{\a}{N}\right),
 \eeq
 where $\a$ is a constant that in principle depends on the other parameters of the model.  
 Here the number of e-folds is restricted to vary in some phenomenologically reasonable finite range. 
  If $f_\a$ is small this indicates that the spectral index follows the power scaling relation in a good approximation. 
 In Figure 1 the function $f_\a$ is evaluated for the initial values that  satisfy all the constraints imposed. The distribution 
 is shown as a function of $n_{\cR\cR}$ in our interval.
\begin{center}
\includegraphics[scale=0.75]{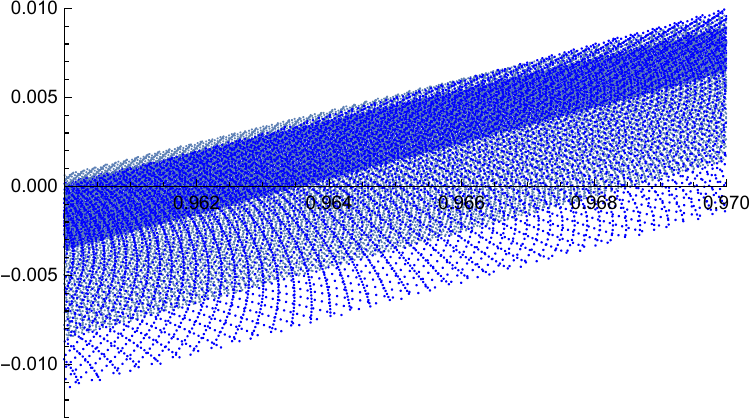}
\end{center}
\begin{quote}
\baselineskip=15pt
{\bf Fig. 1}~An overlay of scaling behavior of the adiabatic spectral index $n_{\cR\cR}$ for the 
energy scales $\mu\in \{35,40,45\}M_\rmPl$.
\end{quote}

\baselineskip=21pt

The distribution of the initial values for the scaling function $f_\a$ associated to the spectral index depends on the 
choice of the energy parameter $\mu$. In particular the width of the distribution varies with $\mu$, becoming smaller 
as the parameter $\mu$ increases.

\subsection{The scaling behavior of  tensor-to-scalar ratio $r(N)$}

 In the singlefield classes briefly highlighted in Section 2 the scaling behavior of the tensor ratio $r(N)$ is described by a power law, 
 hence it is natural to ask whether in multifield inflation  a behavior of the type 
\beq
 r_p(N) ~=~ \frac{\a_p}{N^{\b_p}}
 \eeq 
 for some constants $\a_p, \b_p$
can provide a good description when the theory parameters are held fixed. It is however not clear a priori that 
in multifield inflation the power law is the most appropriate scaling type, even if scaling exists. To the best of our 
knowledge no such analyses have been performed and we adopt an agnostic view about what kind of scaling 
one might consider. 

A first signal that a scaling relation between $r$ and $N$ exists in modular inflation 
 has emerged  in the analysis of \cite{rs21}. There  the distribution of the three 
variables $(n_{\cR\cR}, r, N)$ was considered and the results took the form of two dimensional sheets that were essentially 
degenerate along the direction of the spectral index. As the energy scale $\mu$ is varied, the degeneracy remains 
while the location of the sheets changes 
in this 3D parameter space. An illustration of this effect is shown in Figure 2. It becomes clear from this graph that as $\mu$ varies the precise form of scaling law changes as the sheets move through this 3D-space, i.e. the scaling exponents are energy dependent.
\begin{center}
\includegraphics[scale=0.7]{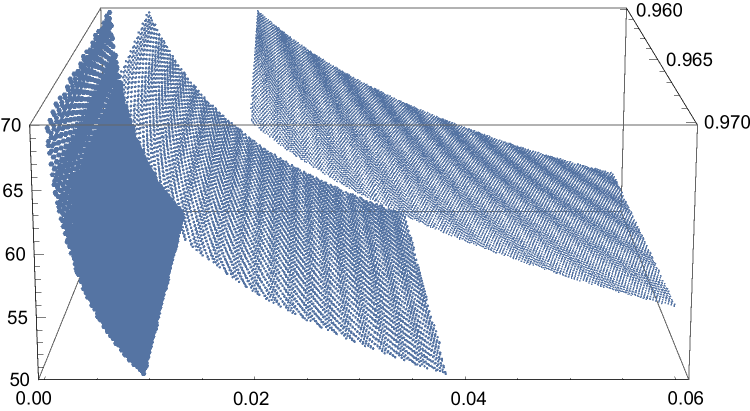}
\end{center}
\begin{quote}
\baselineskip=15pt
{\bf Fig. 2}~Indication of scaling behavior of the tensor-to-scalar ratio by the surfaces swept out by the initial value data 
in the 3D space spanned by the parameters $(n_{\cR\cR}, r, N)$. Here $n_{\cR\cR}$ is centered around 0.965, the tensor 
ratio is bounded by 0.04 and the number of e-folds is chosen to be in the interval $[50, 70]$. 
 The three surfaces are associated to the energy scale
  $\mu/M_\rmPl \in \{30, 35, 40\}$, from left to right.
\end{quote}

\baselineskip=21pt 

The observation illustrated in Figure 2 raises the question what precisely the scaling law is that describes these 
2D distributions in the 3D space. As noted above, the paradigmatic  classes encountered in singlefield inflation 
lead to power laws, hence these provide a natural starting point, even though there is a priori no reason why 
power laws should persist for more general classes of models. 
 It turns out that while the scaling $r = \a_p N^{-\b_p}$ does provide a good approximation to the data, 
 they are not the best functions to describe the tensor ratio $r(N)$. A small but systematic improvement of 
the scaling approximation is obtained by an exponential fit. This is illustrated by a typical example of the $(r,N)$ data distribution
and the two scaling laws for a fixed $\mu$. In Figure 3 we show the scaling distribution and the scaling relations for two representative 
$j-$ and $h_2-$inflation parameters.
\begin{center}
\includegraphics[height=1.7in,width=3in]{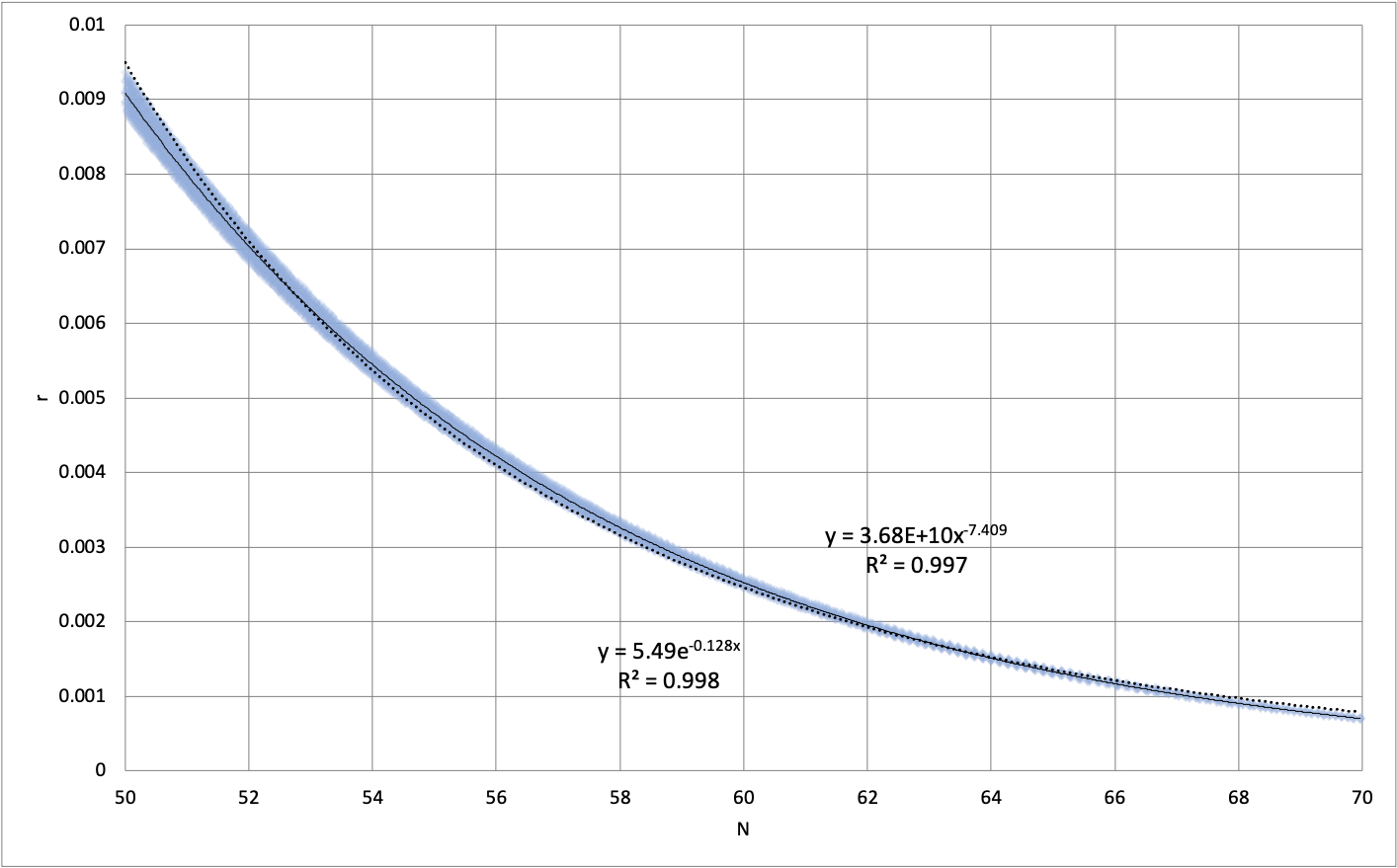}
~~~
\includegraphics[height=1.7in,width=3in]{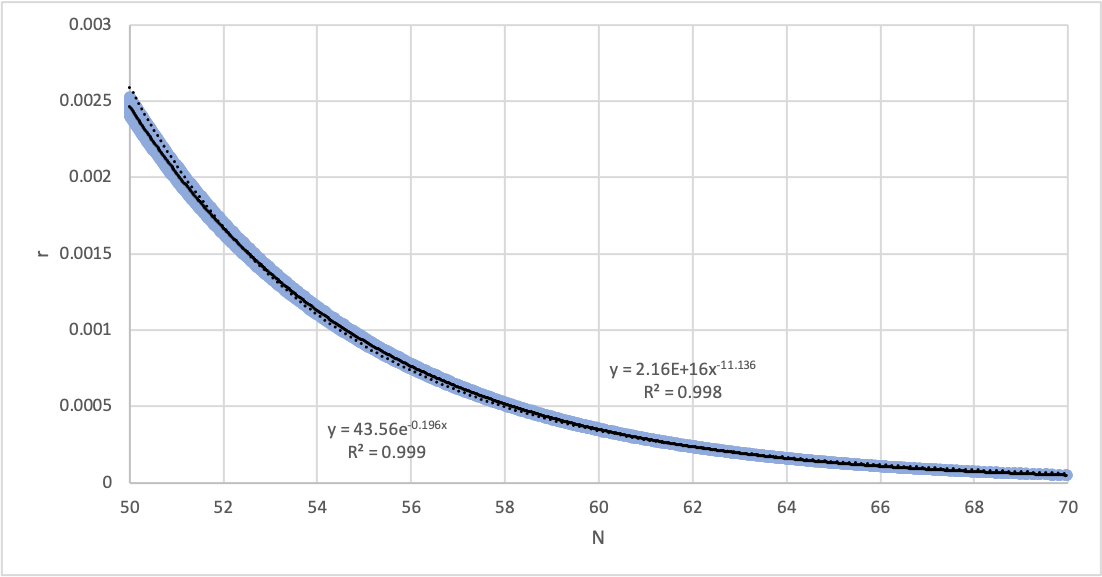}
\end{center}
\baselineskip=15pt
\begin{quote}
{\bf Fig. 3} ~The left panel shows the $(r,N)$ distribution for $j$-inflation at $\mu=30M_\rmPl$ in combination with scaling laws, 
while the right panels shows the same for $h_2$-inflation at $\mu=14M_\rmPl$.
\end{quote}
 
 \baselineskip=21pt
 
The quality of a fit is usually characterized by the $R^2$-values. This parameter contains a denominator that is determined by the 
spread of the data distributions and therefore drives $R^2$ to one for wider distributions. It is useful to illustrate the differences between 
the power ansatz and the exponential ansatz by considering more directly the differences between the data distribution and the fit function. 
We illustrate this in Figure 4 by considering the function
\beq
g_*(N) = r(N)-r_*(N)
\lleq{function-g}
 for $*=p,e$ for the  power and the exponential fit.
 \begin{center}
 \includegraphics[scale=0.22]{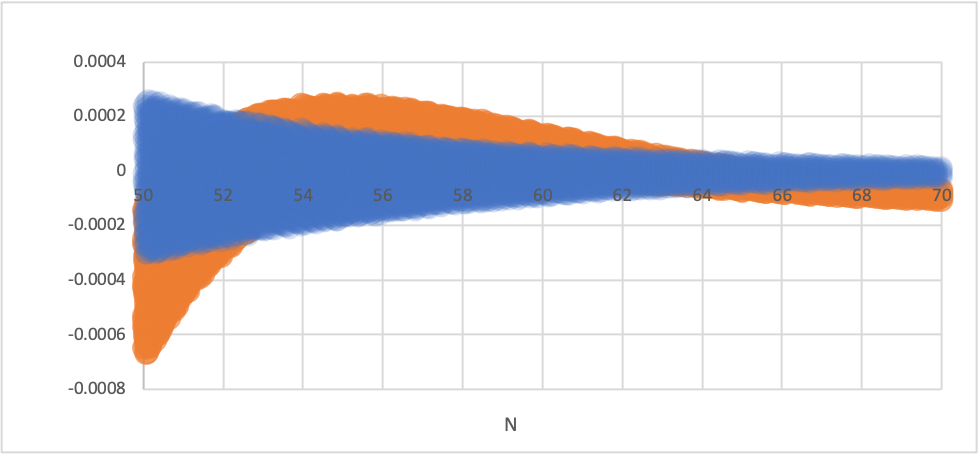}
~~~
\includegraphics[scale=0.22]{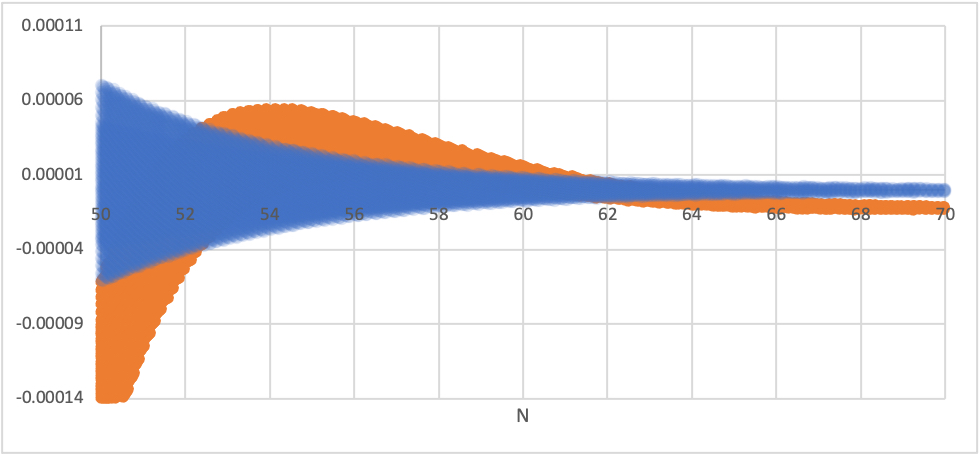}
 \end{center}
 \baselineskip=15pt
 \begin{quote}
 {\bf Fig. 4} ~Distribution of points of eq. (\ref{function-g}) for power scaling $r_p(N)$ (orange) 
 and exponential scaling $r_e(N)$ (blue).
 The left panel shows the function for $j$-inflation at $\mu=30M_\rmPl$, while the right panel shows the result for $h_2$-inflation at 
 $\mu=14 M_\rmPl$.
 \end{quote}
 
 \baselineskip=21pt
 
Varying the energy scale $\mu$ leads to a range of scaling amplitudes and exponents for the $r(N)$ scaling, as illustrated in 
Figure 5, which contains the distribution of
initial values $\tau_{\rm in}$ that satisfy all the imposed observational constraints. Superimposed is the scaling behavior of 
the data obtained by adopting an exponential 
scaling relation of the type
\beq
 r_e(N)  ~=~ \a_e \exp(-\b_e N),
 \eeq
 where $\a_e$ and $\b_e$ are parameters that vary with $\mu$. The variation of $\mu$ produces a
  fan-like family of  fixed-$\mu$ scaling relations parametrized by the amplitudes $\a_e(\mu)$ and the 
scaling exponents $\b_e(\mu)$. This raises the question how these scaling exponents compare to those of other 
theories, which we consider in the next subsection.

\begin{center}
\includegraphics[scale=0.6]{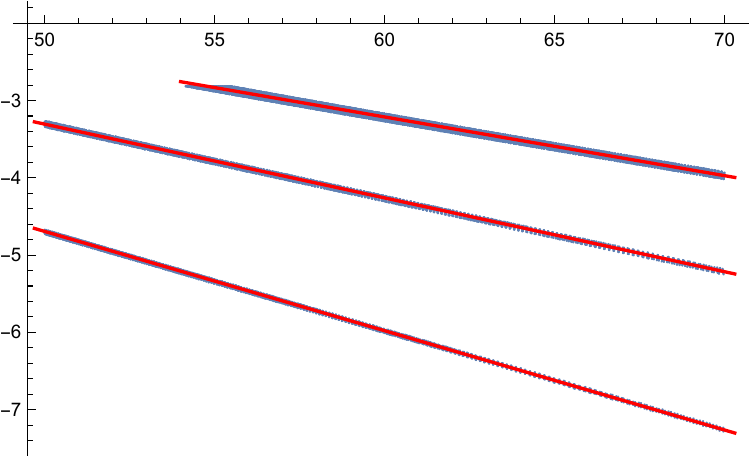}
\end{center}
 \begin{quote}
 {\bf Fig. 5} Exponential scaling behavior of $r(N)$ on a $\log_{10}$ scale 
 for the energy scales  $\mu\in \{30, 35,40\}M_\rmPl$ from bottom to top.
   \end{quote}

\subsection{Comparison with singlefield  scaling classes}

We have shown above that to a  good approximation the data for $j$-inflation and $h_2$-inflation can be described by 
either a power or an exponential scaling ansatz for the tensor-to-scalar ratio, with a preference for the exponential 
scaling.  The power scaling behavior of the tensor ratio $r(N)$ in our models can be compared 
with the  power law scaling behavior of singlefield inflation.   
As described briefly in Section 2, the paradigmatic classes of monomial inflation and Starobinsky type inflation lead to 
 scaling exponents $\b=1$ and $\b=2$ respectively, while the viable hilltop models push this limit to at most  four. 
  These can be compared with the family of exponents of the power scaling relation  $\b_p(\mu)$  in dependence of $\mu$.
  The computations described in this paper show that  the tensor exponents $\b_p(\mu)$ for 
 $j$-inflation have a lower bound of about four but are in general significantly higher, while the tensor exponents for 
 $h_2$ inflation are larger than seven. It follows that the scaling classes of $j$-inflation and $h_2$ inflation 
 are different from either of the singlefield classes
discussed above, thereby defining a different scaling regime.

\vskip .3truein

\section{Scaling behavior of the scaling exponents}

In any theory that exhibits scaling behavior and has more than one parameter
 the scaling exponents $\b$ in general depend on these additional ingredients of the theory. 
 This family of exponents can be used to characterize the model, hence provides an interesting tool for their  classification. 
The new feature introduced by the existence of a family $\b(\mu)$ is that we can in turn analyze the behavior
of these exponents as a function of $\mu$ and test whether the exponents in turn lead to a new scaling relation
 in dependence of $\mu$, leading to an iterative scheme of scaling of scaling exponents. 
 If such an iterative scheme does lead to a new exponent the theory could in principle be characterized 
by a single scaling relation  instead of a family of exponents.

We now apply the idea of an iterative scaling scheme to modular inflation.
We have shown above that for fixed $\mu$ the scaling of the tensor-to-scalar ratios $r(N,\mu)$ of our 
modular inflation theories can be modeled  to a good degree by either 
 a power law, leading to scaling exponents $\b_p(\mu)$, or by an exponential relation with exponents 
 $\b_e(\mu)$, and that the  exponential relation systematically provides a better fit.  
 The variation of $\mu$  thus leads to two families of scaling exponents that can 
 be used to characterize  the models.  
The scaling-of-scaling idea then raises the question whether the $\b_e(\mu)$ 
 show a specific behavior which then would lead to a single parameter for each of the models as a main 
characteristic. In the following we focus on the exponential scaling exponents $\b_e(\mu)$ and 
show that these follow a power law 
\beq
 \b_e(\mu) ~=~ \g~\mu^{-\delta}
 \eeq
with amplitude $\g$ and exponent $\delta$. The models are then characterized  by a single pair $(\g, \delta)$.

In the analysis discussed here the energy scale $\mu$ is varied over
 the range $\mu\in [16, 45]M_\rmPl$ for $j$-inflation, and the range $\mu \in [12, 18]$
for $h_2$-inflation. In both cases we find that  a power law model of the family of  
exponents $\b_e(\mu)$  provides a better fit than an exponential model. This is 
illustrated in Figure 6.
\begin{center}
\includegraphics[width=3.2in, height=1.8in]{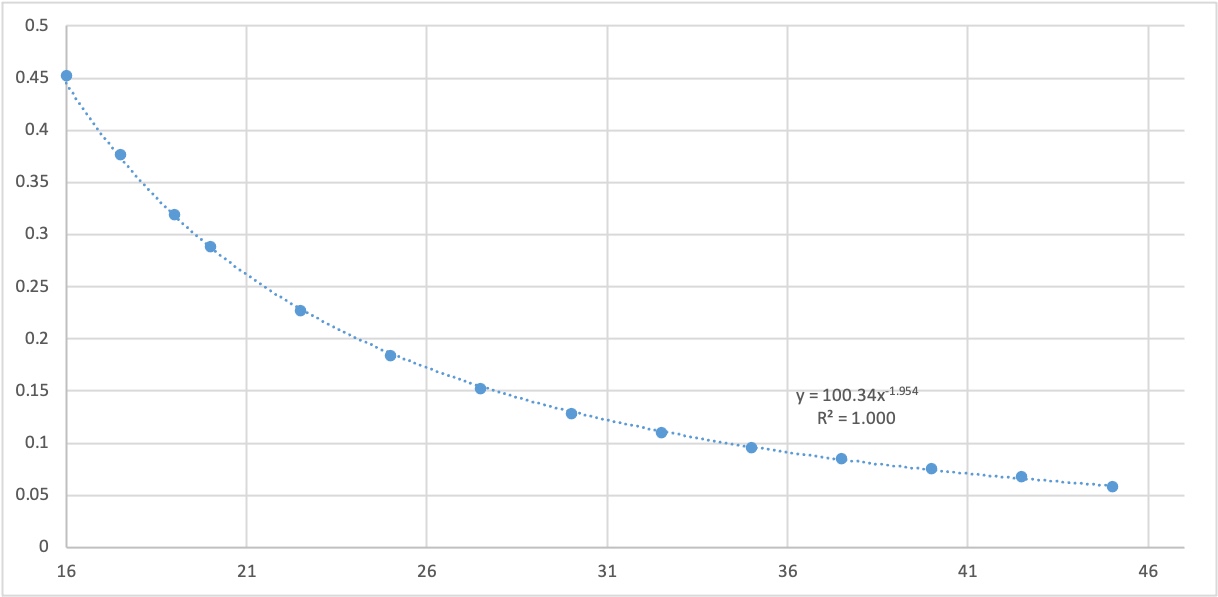}
~~~
\includegraphics[width=3in, height=1.8in]{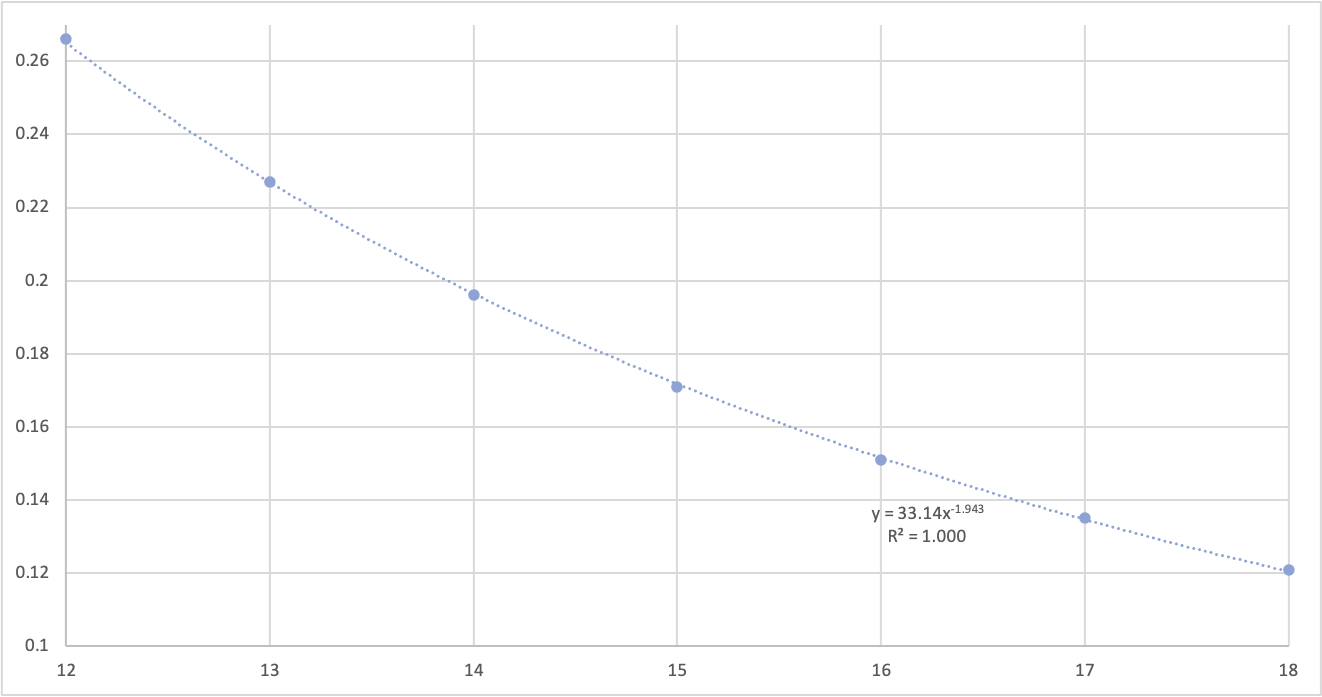}
 \end{center}
 \begin{quote}
 {\bf Fig. 6} The scaling behavior of the exponents $\b_e$ (vertical axis) as a function of $\mu$ (horizontal axis) 
  for $j$-inflation in the left panel and $h_2$ inflation  in the right panel.
 \end{quote}

 The unexpected result here is that the exponents $\delta$ for $j$-inflation and $h_2$-inflation agree quite 
 closely even though the potentials 
 are different and the range of the energy scale is different in these models, and so are the 
scaling exponents $\b_e(\mu)$. This agreement for these two exponents $\delta^j, \delta^{h_2}$ is valid 
at the per-mille level $(\delta^j - \delta^{h_2})/\delta^j  \cong 10^{-3}$.
 We will comment on this in the next section.

 The question arises  whether the energy dependent scaling behavior found here 
 could be understood analytically, if only in retrospect. The difficulty in doing so is that it is not possible to invert the 
 e-fold formula analytically.   This problem is encountered already in singlefield inflation models with not 
 very complicated potentials, where it is also not possible to obtain 
  energy dependent scaling exponents  analytically if only the slow-roll approximation is assumed. The scaling relations mentioned 
 in section two are obtained only after imposing approximations that go beyond the slow-roll approximations and result 
 in scaling exponents that are energy independent. Hence the scaling behavior trivializes in this limit.
In modular inflation one encounters the additional problem that all the observables are complicated functions 
of two variables.

\section{Conclusion}

In this paper we have shown that even theories with complicated potentials for inflaton vectors can exhibit 
scaling behavior of their observables. We have taken these results as the starting point for a classification 
scheme within the general framework of multifield inflation in terms of the 
scaling behavior of their basic observables, in particular the tensor ratio as a function of the number of e-folds. 
The question whether a correlation exists between the tensor ratio $r$ and the number of e-folds can be 
asked for any inflationary model, independent of the structure of the potentials. Hence this approach is based
 on a universal structure present in all inflation models and can in principle be applied to any theory. While in some 
singlefield cases the scaling behavior can be made explicit analytically, given appropriate approximations, in more general 
cases such scaling can only be investigated numerically.  Historically, the focus in the literature has been on 
power law scaling for both the spectral index and the tensor ratio with exponents that have been used to 
identify certain universality classes, motivated by the two oldest classes of models, monomial and Starobinsky 
type inflation.  In this paper our focus has been on inflationary models with more than one inflaton component,
exemplified here by the framework of modular inflation.
We have seen that power scaling does give a good description of the data. In this approximation 
a direct comparison with the singlefield scaling exponents discussed in the literature is possible, with the result 
that modular inflation has quite different scaling exponents, thereby leading to a different scaling regime compared 
to the singlefield models considered in section two.
From a phenomenological perspective, the 
refinement considered here of energy dependent scaling exponents should eventually lead, with improved data 
from either the CMB or large scale structure collaborations, to the possibility of constraining the parameter dependence 
in a new way.

In any theory with more than one parameter the CMB results constrain, but in general do not determine,
these parameters. Hence, if scaling exists in a model the exponents will vary with these parameters, leading 
to exponent functions $\b_e(\mu)$. We can then consider an iterated scaling scheme in which we ask 
whether the models lead to  a scaling of scaling exponents phenomenon. 
We have shown that in the two modular inflation models considered here such an iteration does exist and we have 
seen that it can be described by a power scaling relation, leading to a single exponent $\delta$ that 
characterizes each of the modular inflation models. 
The usefulness of this new parameter is indicated by the fact that although the exponents $\b_e(\mu)$ are different 
for $j$-inflation and $h_2$-inflation, the parameters $\delta^j$ and $\delta^{h_2}$ are very close. Hence $\delta$ 
provides a characteristic that is shared by the two modular inflation models analyzed here.
In general our proposed diagnostic tool given by the functional behavior of the energy dependent 
scaling exponents will be different for different models, thereby providing a further discriminating function
 on the space of all inflationary models. 
This can be seen already by comparing  our modular inflation results with the energy independent 
scaling relations reviewed in section two.

 It would be interesting to explore whether energy dependent scaling relations exist in other inflationary models, 
  either with a single component or with multicomponent fields, 
  such as hybrid inflation models \cite{l93, c20etal}, spectator models such as in ref. \cite{kt20},
other hyperbolic models, such as those considered in \cite{b17, mm17, m19etal}, 
 or the types of models considered in refs. \cite{ls17, bl17, bl18, a19etal, bl22}, or \cite{pr18, apr19}. 
If scaling does exist in other models then it would be interesting to see how the corresponding exponents  
relate to the scaling behavior described here. 

\vskip .3truein

{\bf Acknowledgement.} \\
It is a pleasure to thank Burt Carter,  Ali Ishaq, Jack Morse and Vipul Periwal for discussions and correspondence.

 \end{document}